\documentclass[pra,twocolumn,showpacs]{revtex4}
\usepackage{graphicx}

\begin{document}

\title{A theoretical scheme of thermal-light ghost imaging by $N$th-order intensity correlation}
\author{Ying-Chuan Liu and Le-Man Kuang\footnote{Author to whom any correspondence should be
addressed. }} \affiliation{Key Laboratory of Low-Dimensional
Quantum Structures and Quantum Control of Ministry of Education,
and Department of Physics, Hunan Normal University, Changsha
410081, People's Republic of China}

\begin{abstract}
In this paper, we propose a  theoretical scheme of ghost imaging
in terms of $N$th-order correlated thermal light. We obtain the
Gaussian thin lens equations in the ghost imaging protocol. We
show that it is possible to produce $N-1$ ghost images of an
object at different places in a nonlocal fashion by means of a
higher-order correlated imaging process with an $N$th-order
correlated thermal source and correlation measurement. We
investigate the visibility of the ghost images in the scheme, and
obtain the upper bounds of the visibility for the $N$th-order
correlated thermal-light ghost imaging. It is found that the
visibility of the ghost images can be dramatically enhanced when
the order of correlation becomes larger.
\end{abstract}
 \maketitle

\section{Introduction}

Ghost imaging with thermal light
\cite{ben,gat1,gat2,gat3,che,ca,cli,bra,val,fer,zh,mey,er,gat4,bac,cai,zha,scar,dan,liu,zhan}
has been studied extensively in recent years. Bennink and
coworkers \cite{ben}  first pointed out that ghost imaging can
also be realized using a classical source with the appropriate
correlations. A thermal or quasi-thermal source can exhibit such a
classical correlation. A very close formal analogy was
demonstrated between ghost imaging with thermal and
quantum-entangled beams in Refs. \cite{gat2,gat3,che,ca}, which
implied that classically correlated beams were able to emulate the
relevant features of quantum ghost imaging. Number of experiments
on thermal light ghost imaging have been performed in Refs.
\cite{val,fer,zh,mey,zhan}. A unified treatment of classical and
quantum ghost imaging was established in terms of Gaussian-state
analysis in Ref. \cite{er}.

Recently, some attention has been paid to the physics of
thermal-light ghost imaging
\cite{mey,shi,erk,ga,sca,wa,bre,sh,sha} and higher-order-coherence
or correlation effects of thermal light
\cite{ou,ba,li,dz,aga,ric}. The higher-order coherence or
correlation has shown attractive properties in practical
applications. Multiphoton imaging with thermal light is one of
these exciting areas. In a previous work, our group theoretically
initiated a study of thermal-light ghost imaging in terms of
higher-order correlated thermal light \cite{ou}. We proposed a
thermal-light ghost imaging scheme with third-order correlated
thermal light. In this scheme, a third-order correlated
thermal-light source, a test optical arm and two reference optical
arms are used to produce two ghost images. Two ghost images are
created at two different places in a nonlocal fashion as a
consequence of the third order correlation of the involved optical
fields. It was shown that the third-order correlated imaging
includes richer correlated imaging effects than the second-order
correlated one. In this paper, we want to propose a thermal-light
ghost imaging scheme with $N$th-order intensity correlation and
investigate the visibility of the ghost images. We show that the
usual second- and third-order ghost imaging are only two
particular examples of our present scheme.

The paper is organized as follows.   In Sec. II,  begin with the
thermal-light source with $N$th-order correlation, we will propose
the thermal-light ghost imaging scheme with $N$th-order intensity
correlation. In Sec. III, the visibility of the ghost images in
the proposed scheme will be investigated.  Finally, we shall
conclude our paper with discussions and remarks in the last
section.

\section{Ghost imaging scheme with $N$-order correlated thermal light}

The basic setup for ghost imaging with $N$-order correlated
thermal light is indicated in Figure 1 which includes one test
optical arm and $N-1$ reference optical arms. First arm is the
test arm. An unknown object with transmission function $T(x)$ and
a collective lens with focal length $f_c$ are placed on the test
arm while there is one imaging lens with focal lengths $f_i$ on
each reference arm. The object is placed at the focal plane of the
collective lens at distance $z_1$ from the thermal-light source.
The distance between $i$th imaging lens and the thermal-light
source is $z_{i0}$ with $i=2,3, \cdots, N$. The first detector
$\tt{D_1}$ is a bucket detector placed at the focal plane of the
collective lens on the right-hand side. The other detectors are
scanning detectors $\tt{D_i}$ with $i=2,3, \cdots, N$ which  are
placed at distances $z_{i1}$ from $i$th imaging lens. The signals
from the $N$ photon counting detectors are sent to an electronic
coincidence circuit to measure the rate of coincidence counts.

The thermal light source with $N$th-order intensity correlation,
usually obtained by illuminating a laser beam into a slowly
rotating ground glass, is divided into $N$ beams, which can be
implemented by an appropriate combination of $N-1$ beam splitters.
Consider a monochromatic plane wave described by the field
$E_{0}\exp[i(k_{0}z-w_{0}t)]$ illuminating a material containing
disordered scattering centers. After scattering, the field can be
written as $E(x,z,t)=\int E(\mathbf{q})\exp[i(q \cdot x
+k_{z}z-w_{0}t)]d\mathbf{q}$ where $\mathbf{q}$ is the transverse
wave vector introduced by the random scattering and satisfies the
relation $|\mathbf{q}|^{2}+k_z^{2}=k_0^{2}$. Hence,
$E(\mathbf{q})$ is a stochastic variable obeying Gaussian
statistics. However, the scattered waves with different transverse
wave vectors are statistically independent. If $|\mathbf{q}|\ll
k_0$, the scattered field can be approximately written as
$E(x,z,t)=A(x)\exp[(i(k_{0}z-w_{0}t)]$ where $A(x)=\int E(
\mathbf{q})\exp[i(\mathbf{q }\cdot \mathbf{x} )]d\mathbf{q}$ is
the slowly varying envelope. As a result, we have defined a
monochromatic thermal light random in both strength and
propagation direction. According to the Wiener-Khintchine theorem,
the first-order spectral correlation satisfies the following
expression
\begin{eqnarray}
\label{1}
 \langle E^ \ast (\mathbf{q})
 E(\mathbf{q'})\rangle=S(\mathbf{q}) \delta (\mathbf{q}-\mathbf{q'}),
\end{eqnarray}
where $S(\mathbf{q})$ is the power spectrum of the spatial
frequency. For any field with thermal statistics, all high-order
correlations can be expressed in terms of the first-order ones due
to $\langle E(\mathbf{q})\rangle=0$ \cite{man}. Then, $N$th-order
spectral correlation of thermal light can be written as
\begin{eqnarray}
\label{2} \left\langle\prod^N_{i=1} E^{\ast
}(\mathbf{q}_{i})E(\mathbf{q}_{i}^{\prime })\right\rangle &=&
\sum_{\begin{array}{c} r_i\neq r_j\\
s_i\neq s_j\end{array}}^N  \left(\prod^4_{i=1}\delta
(\mathbf{q}_{r_i}-\mathbf{q}_{s_i}^{\prime })\right)'\nonumber\\
&&\times S(\mathbf{q}_{1})S(\mathbf{q}_{2})\cdots
S(\mathbf{q}_{N}),
\end{eqnarray}
where the prime in the summation means that all of repeating terms
are subtracted from the summation over $r_i$ and $r_j$. After
passing through a combination of $N-1$ beam splitters, the thermal
light with $N$th-order spectral correlation is divided into $N$
correlated thermal-light beams which are input light beams of the
test and reference arms. For simplicity, we consider the
one-dimensional case, and let $x_{0}$ and $x_{n}$ with $n = 1,
2,\cdots, N$ be the transverse coordinates of the source plane and
detection planes, respectively. Here we have assumed that the
starting planes of the $N$ thermal-light beams are overlapped. Let
$H_{1}(x_{1}, x_{0})$ be the impulse response function of the test
arm,  and $H_{n}(x_{n}, x_{0})$ with $n = 2, 3,\cdots, N$  be the
impulse response function of $n$th reference arm, respectively.
Assume that the light field on the $n$th detection planes is
denoted as $E(x_{n})$, which is connected with the optical field
of the thermal-light source through the following relation
\begin{equation}\label{3}
E\left( x_{n}\right) =\int h_{n}\left( x_{n},-q_n\right) E\left(
q_n\right) dq_n
\end{equation}
where $h_{n}(x_{n},q_n)=(1/\sqrt{2\pi})\int H_{n}(x_{n},
x_{0})\exp(-iq_nx_{0})dx_{0}$ is the Fourier transformation of the
impulse response function $H_{n}(x_{n}, x_{0})$. Then, the
$N$th-order correlation function of the joint intensity  at the
$N$ detection planes can be expressed as
\begin{eqnarray}\label{4}
&&G^{\left( N\right) }\left( x_{1},x_{2},\cdots,x_{N}\right)
=\left\langle \prod^N_{i=1}E^{\ast }\left( x_{i}\right) E\left(
x_{i}^{\prime }\right) \right\rangle.
\end{eqnarray}

The rate of coincidence counts is governed by the $N$th-order
correlation function given by equation (\ref{4}) which can be
calculated in terms of the impulse response functions of the
relevant optical systems. The impulse response functions
\cite{Cao} in the reference arms can be written as
\begin{eqnarray}\label{5}
h_{r}(x_{r},q)&=&\sqrt{\frac{f_{r}}{2\pi \left(f_{r}-z_{r1}\right)}}e^{i\varphi_{r}(x_{r},q)},\\
\label{6} \varphi_{r}(x_{r},q)&=&k\left( z_{r0}+z_{r1}\right)
-\frac{q^{2}}{2k}\left(z_{r0}+\frac{z_{r1}f_{r}}{f_{r}-z_{r1}}\right)
\nonumber \\
&&-\frac{(2qf_{r}+kx_{r})x_r}{2(f_{r}-z_{r1})},
\end{eqnarray}
where $z_{r1}\neq f_{r}$ with $r=2,3,\cdots, N$, and the impulse
response function in the test arm is given by
\begin{eqnarray}\label{7}
h_{1}(x_{1},q)&=&\frac{1}{2\pi }\sqrt{\frac{k}{if_{c}}}\exp \left[
ik\left( z_{1}+2f_{c}\right) +\left(
-i\frac{z_{1}q^{2}}{2k}\right)
\right]\nonumber \\
&&\times\int T\left( x\right) \exp \left[ -i\left(
\frac{kx_{1}}{f_{c}}+q\right) x\right] dx.
\end{eqnarray}

\begin{figure}[htp] \center
\includegraphics[scale=0.6]{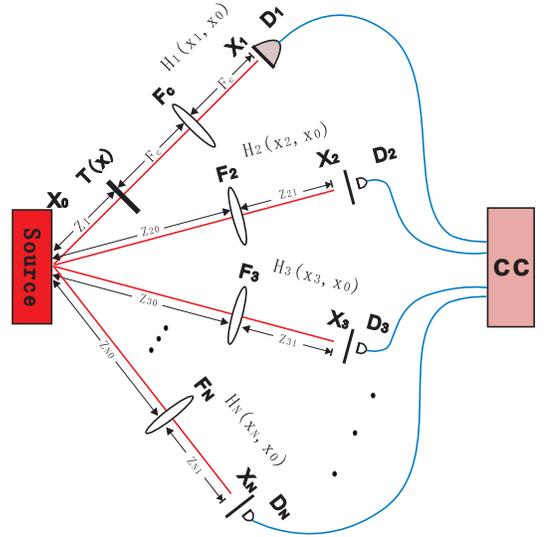}
\caption{(Color online) Schematic of the setup for implementing
ghost imaging with $N$th-order correlated thermal light. It
consists of one test arm and $N-1$ reference arms. The test arm
includes the object $T(x)$, the collection lens $\tt{F_{c}}$, and
the bucket detector  $\tt{D_1}$. The object and the detector are
placed in its two focal planes of the collection lens,
respectively. Each reference arm includes one imaging lens and a
scanning detector denoted by $\tt{F_{i}}$ and  $\tt{D_i}$ with
$i=2, 3, \cdots, N-1$, respectively.}
\end{figure}

In order to obtain the $N$th-order correlation function given by
equation (4), we first calculate the following integrations:
\begin{eqnarray}\label{8}
I_{r} &=&\int S\left( q\right) \left\vert h_{r}\left(
x_{r},-q\right) \right\vert ^{2}dq, \\
\label{9} C_{rr^{\prime }} &=&\int S\left( q\right) h_{r}^{\ast
}\left( x_{r},-q\right) h_{r^{\prime }}\left( x_{r^{\prime
}},-q\right)dq,
\end{eqnarray}
where both $r$ and $r'$ may take $1,2,\cdots, N$,  but $r\neq r'$.
In the broadband limit, S(q) can be regarded as a constant S(0).
Hence we have the total intensity of the thermal light $S_{0}=\int
S(q)dq\approx S(0)q_{0}$, where $q_{0}$ is the spectral bandwidth
of the source. Substituting the impulse response functions of the
test and reference arms given by equations (\ref{5}) and
(\ref{7}) into equations (\ref{8}) and  (\ref{9}), we obtain
\begin{eqnarray}\label{10}
I_{1} &=&\frac{S(0)k}{4\pi ^{2}f_{c}}\int \left\vert T\left(
x\right) \right\vert ^{2}dx, \hspace{0.3cm}
I_{r} =\frac{f_{r}S(0)q_{0}}{2\pi \left( f_{r}-z_{r1}\right) }, \\
\label{11} C_{r1}&=&\frac{S(0)}{2\pi }\sqrt{\frac{kf_{r}}{i2\pi
f_{c}\left( f_{r}-z_{r1}\right) }}\int T(x)e^{i\phi_{r
1}(x,q)}dxdq, \nonumber
\\
\\
\label{12}C_{rr^{\prime }} &=&\frac{S(0)}{2\pi }\sqrt{\frac{f_{r}f_{r^{\prime }}}{%
\left( f_{r}-z_{r1}\right) \left( f_{r^{\prime }}-z_{r^{\prime
}1}\right) }}\int  e^{i\phi_{rr'}(q)}dq,
\nonumber \\
\end{eqnarray}
where both $r$ and $r'$ may take $2,3,\cdots, N$   but $r\neq r'$,
and we have introduced two phase functions
\begin{eqnarray}
\label{13}
\phi_{r 1}(x,q) &=&\frac{q^2}{2k}\left(z_{r0}-z_1+\frac{z_{r1}f_r}{f_r-z_{r1}}\right)-\left(q+\frac{kx_1}{f_c}\right)x \nonumber\\
& &+ qX_r -k(z_{r0}+z_{r1}-z_1-2f_c),\\
\label{14} \phi_{rr'}(q) &=&\frac{kx^2_{r}}{2\left(
f_{r}-z_{r1}\right) }-\frac{kx^2_{r'}}{2\left( f_{r^{\prime
}}-z_{r^{\prime
}1}\right)} \nonumber\\
&&-\frac{q^2}{2k}\left(z_{r'0}-z_{r0}+\frac{z_{r'1}f_r'}{f_r'-z_{r'1}}-\frac{z_{r1}f_r}{f_r-z_{r1}}\right)\nonumber\\
&&-q\left(X_{r'}-X_r\right)+k(z_{r'0}+z_{r'1}-z_{r0}-z_{r1}),\nonumber \\
\end{eqnarray}
where  the scaled positions  $X_r=f_rx_r/(f_r-z_{r1})$.

From equations (11) and (13), it is straightforward to see that
when the positions of the object, ghost images, and the lenses
obey the following Gaussian thin lens equations for the
$N$th-order correlated imaging
\begin{equation}\label{15}
\frac{1}{z_{r0}-z_{1}}+\frac{1}{z_{r1}}=\frac{1}{f_{r}}\text{ \ \
\ \ \ \ } (r=2,3,\cdots,N),
\end{equation}
the cross-correlation functions of intensity fluctuation can be
simplified as
\begin{eqnarray}\label{16}
C_{r1}&=&\frac{S(0)}{2\pi }\sqrt{\frac{kf_{r}}{i2\pi f_{c}\left(
f_{r}-z_{r1}\right) }}\exp\left(-\frac{ikx_1X_r}{f_c}\right)
T(X_r), \nonumber
\\
\\
\label{17}C_{rr^{\prime }} &=&\frac{S(0)}{2\pi }\sqrt{\frac{f_{r}f_{r^{\prime }}}{%
\left( f_{r}-z_{r1}\right) \left( f_{r^{\prime }}-z_{r^{\prime
}1}\right) }}e^{i\phi_{rr'}(0)}\delta(X_{r'}-X_r).
\nonumber \\
\end{eqnarray}

When Gaussian thin lens equations are satisfied, and  $X_2\neq
X_3\cdots\neq X_N$ the $N$th-order correlation function of the
joint intensity  at the $N$ detection planes given by equation
(\ref{4}) can be reduced to the following form
\begin{eqnarray}\label{18}
&&G^{\left( N\right) }\left( x_{1},x_{2}, \cdots, x_{N}\right)
\nonumber \\
&&= \sum^N_{r_i\neq r_j=2}\left(I_{r_1}I_{r_2}\cdots
I_{r_{N-2}}|C_{r_{N-1} 1}|^2\right)'+I_1 I_2\cdots I_N,
\nonumber\\
\end{eqnarray}
where the prime on the right-hand side of above equation means
that repeating terms in the summation over $r_i$ and $r_j$ should
be subtracted. The last term on the right-hand equation (18) is
the background term which is the multiplication of the intensity
distribution at the $N$ detectors. It does not contribute to the
correlated imaging, but it may affect the visibility of produced
ghost images. Each term in the summation over $r_i$ and $r_j$ is
the multiplication of the intensity distribution at detector
$\tt{D_i}$ and the intensity fluctuation correlation between
$\tt{D_1}$ and $\tt{D_i}$ with $i=2,3,\cdots, N$, which gives the
information of the object imaged at $\tt{D_i}$.

In particular, when the $N-1$ reference arms are identic, i.e.,
$f_{2}=f_{3}=\cdots=f_{N}$, $z_{20}=z_{30}=\cdots=z_{N0}$ and
$z_{21}=z_{31}=\cdots=z_{N1}$, we have $I_2=I_3=\cdots=I_N\equiv
I$, $N$th-order correlation function becomes
\begin{equation}\label{19}
G^{(N)}=I^{N-1}I_{1}+\chi I^{N-2}\sum^N_{i=2}|T(X_i)|^2,
\end{equation}
where we have introduced the parameter $\chi=S^{2}(0)kf_{r}/[8\pi
^{3}f_{c}\left( f_{r}-z_{r1}\right)]$.

Equation (19) indicates that for the object placed at the test
arm, $N-1$ ghost images can be produced in the $N-1$ reference
arms through coincidence count measurements upon $N$ detectors.
Each ghost image is an amplified image of the object with the
amplifying rate $f_{r}/(f_{r}-z_{r1})$ with $r=2,3,\cdots, N$
since $X_r=f_rx_r/(f_r-z_{r1})$. From the Gaussian thin lens
equations of  $N$th-order correlated imaging given by equation
(15) we can see that the higher-order correlated imaging exhibits
richer imaging effects. In fact, the imaging equation (15)
indicates that $z_{r0}-z_1 $ is the object distance while $z_{r1}$
is the image distance for $r$th joint path with $r=2, 3,\cdots,
N$. Just as the ordinary imaging law, when the object distance of
the $r$th joint path $z_{r0}-z_1 $ is greater (less) than the
focal length $f_r$, the correlated image is real (virtual). As a
specific example, we consider the case of $N=4$. In this case,
from the imaging equation (15) we can obtain the following eight
ghost-image configurations: (1) when $z_{r0}-z_1>f_r$ with
$r=2,3,$ and $4$, the three ghost images are real; (2) when
$z_{r0}-z_1<f_r$ with $r=2,3,$ and $4$, the three ghost images are
virtual; (3) when $z_{20}-z_1> f_2$, $z_{30}-z_1 < f_3$ and
$z_{40}-z_1 < f_4$, one ghost image is real at the position $x_2$,
the other two ghost images are virtual at the positions $x_3$ and
$x_4$; (4) when $z_{30}-z_1> f_3$, $z_{20}-z_1 < f_2$ and
$z_{40}-z_1 < f_4$, the ghost image at the position $x_3$,  two
ghost images at the positions $x_2$ and $x_4$  are virtual; (5)
when $z_{40}-z_1> f_4$, $z_{20}-z_1 < f_2$ and $z_{30}-z_1 < f_3$,
the ghost image at the position $x_4$ is real, two ghost images at
the positions $x_2$ and $x_3$  are virtual; (6) when $z_{20}-z_1>
f_2$, $z_{30}-z_1 < f_3$ and $z_{40}-z_1 < f_4$, two ghost images
at the positions $x_2$ and $x_3$ are real, the ghost image at the
position $x_4$ is virtual; (7) when $z_{20}-z_1> f_2$, $z_{40}-z_1
< f_4$ and $z_{30}-z_1 < f_3$, two ghost images at the positions
$x_2$ and $x_4$ are real,  the ghost image at the position $x_3$
is virtual; (8) when $z_{30}-z_1> f_3$, $z_{40}-z_1 < f_4$ and
$z_{20}-z_1 < f_2$, two ghost images at the positions $x_3$ and
$x_4$ are real,  the ghost image at the position $x_2$ is virtual.

It is obvious  that when $N=2,3$, the usual results of the second-
and third-order thermal-light ghost imaging are recovered.

\section{Visibility of ghost images}

The visibility is an important parameter to estimate the quality
of a ghost image. We first generalize the visibility of the
second-order correlated imaging \cite{Bache,gat4} to the case of
$N$th-order correlated imaging as follows
\begin{equation}\label{20}
V^{\left( N\right) }=\frac{\left[ G^{\left( N\right)
}-\prod^N_{i=1}\langle I_{i}\rangle\right] _{\max } }{\left[
G^{\left(N\right) }\right] _{\max }}.
\end{equation}

For our present scheme of the $N$th order thermal-light imaging,
substituting equations (19) into equation (20), we obtain the
visibility
\begin{equation}\label{21}
V^{\left( N\right)
}=\frac{\left[\sum^N_{i=2}|T(X_i)|^2\right]_{\max}}{\left[q_{0}\int
\left\vert T\left( x\right) \right\vert
^{2}dx+\sum^N_{i=2}|T(X_i)|^2\right]_{\max}},
\end{equation}
which indicates that the visibility depends on not only the
spectral bandwidth of the thermal-light source $q_0$  and the
transmission function $T(x)$ but also the order number of the
correlation of the thermal-light source. The visibility can be
enhanced with the decrease of the spectral bandwidth. An increase
of the transmission leads to a decrease of the visibility since
when the transmission area increases, more points contribute to
the background that directly makes the visibility decrease.

In order to see the relationship between the visibility and the
order number of the correlation, making use of equations (10) and
(11), we have
\begin{equation}\label{22}
\frac{\left\langle I_{1}\right\rangle \left\langle
I_{r}\right\rangle }{\left\vert \left\langle
E^{*}(x_{1})E(x_{r})\right\rangle \right\vert ^{2}}=q_{0}\int
\left\vert T\left( x\right) \right\vert ^{2}dx,
\end{equation}
where $r=2, 3, \cdots, N$. According to the Cauchy-Schwartz
inequality $\left\langle I_{1}\right\rangle \left\langle
I_{r}\right\rangle \geq \left\vert \left\langle E^{\ast }\left(
x_{1}\right) E\left( x_{r}\right) \right\rangle \right\vert ^{2}$,
which means that $q_{0}\int \left\vert T\left( x\right)
\right\vert ^{2}dx \geq 1$. Hence, we find that
\begin{equation}\label{23}
V^{\left( N\right)
}\leq\frac{\left[\sum^N_{i=2}|T(X_i)|^2\right]_{\max}}{\left[1+\sum^N_{i=2}|T(X_i)|^2\right]_{\max}}\leq
\frac{N-1}{N},
\end{equation}
which implies that the visibility of the ghost images can be
dramatically enhanced when the order of correlation becomes
larger. For $N$th order correlated thermal-light imaging, the
upper bounds of the visibility is given by $V^{(N)}_b=(N-1)/N$. As
expected, when $N=2$ the upper bound of the visibility of the
second correlated imaging  \cite{gat4} is $1/2$.

\section{Concluding remarks}

In conclusion,  we have proposed a theoretical scheme of ghost
imaging in terms of $N$th-order correlated thermal light. Our
scheme includes one test arm and $N-1$ references arms. One object
is placed  on the test arm. $N-1$ ghost images are produced in the
reference arms in a nonlocal fashion by means of a higher-order
correlated imaging process with an $N$th-order correlated thermal
source and correlation measurements. We have derived the Gaussian
thin lens equations which the positions of the ghost images obey
in the ghost imaging protocol. We have also investigated the
visibility of the ghost images in the scheme, and obtained the
upper bounds of the visibility for the $N$th-order correlated
thermal-light ghost imaging. It has been shown that the visibility
depends on not only the spectral bandwidth of the thermal-light
source  and the transmission area of the object but also the order
number of the correlation of the thermal-light source. It is found
that the visibility of the ghost imagines can be dramatically
enhanced when the order of correlation becomes larger. Our present
scheme is a many-ghost imaging protocol. On one hand, it gives
rise to a theoretical origin for developing many-ghost imaging
technology. This gives rise to the possibility of experimentally
producing correlated many ghost images. In fact, the higher-order
correlated imaging opens up new avenues for realizing multi-port
information processing. On the other hand, physically these ghost
images stem from higher-order coherence or correlation of optical
fields. In this sense, the appearance of the many ghost images
reveals an observable physical effect of higher-order coherence or
correlation of optical fields. Hence, it is of very significance
to study higher-order correlated imaging not only for well
understanding the essential physics behind the higher coherence or
correlation of optical fields but also for developing multi-port
information processing technology. The experimental realization
for the many ghost imaging protocol proposed here and practical
applications of the  many ghost imaging phenomenon deserves
further investigation.

\acknowledgments This work was supported by the National
Fundamental Research Program Grant No.  2007CB925204, the National
Natural Science Foundation under Grant Nos. 10775048 and 10325523,
and the Education Committee of Hunan Province under Grant No.
08W012.



\begin{thebibliography}{99}
\bibitem {ben}  R. S. Bennink, S. J. Bentley, and R. W. Boyd, Phys. Rev. Lett. \textbf{89}, 113601 (2002).
\bibitem {gat1} A. Gatti, E. Brambilla, and L. A. Lugiato, Phys. Rev. Lett. \textbf{90},133603 (2003).
\bibitem {gat2} A. Gatti, E. Brambilla, M. Bache, and L. A. Lugiato, Phys. Rev. Lett. \textbf{93}, 093602 (2004).
\bibitem {gat3} A. Gatti, E. Brambilla, M. Bache, and L. A. Lugiato, Phys. Rev. A \textbf{70}, 013802 (2004).
\bibitem {che}  J. Cheng and S. S. Han, Phys. Rev. Lett. \textbf{92}, 093903 (2004).
\bibitem {ca}   Y. Cai and S. Y. Zhu, Opt. Lett. \textbf{29}, 2716 (2004).
\bibitem {cli}  K. W. C. Chan, M. N. O'Sullivan, and R. W. Boyd, Phys. Rev. A \textbf{79}, 033808 (2009).
\bibitem {bra}  E. Brambilla, A. Gatti,  M. Bache, and L. A. Lugiato, Fortschr. Phys. \textbf{52}, 1080 (2004).
\bibitem {val}  A. Valencia, G. Scarcelli, M. D'Angelo, and Y. H. Shih, Phys. Rev. Lett. \textbf{94}, 063601 (2005).
\bibitem {fer}  F. Ferri, D. Magatti, A. Gatti, M. Bache, E. Brambilla, and L. A. Lugiato, Phys. Rev. Lett.\textbf{ 94}, 183602 (2005).
\bibitem {zh}   D. Zhang, Y. H. Zhai, L. A. Wu, and X. H. Chen, Opt. Lett. \textbf{30}, 2354 (2005).
\bibitem {mey}  R. Meyers, K. S. Deacon, and Y. H. Shih, Phys, Rev. A \textbf{77}, 041801 (2008).
\bibitem {er}   B. I. Erkmen and J. H. Shapiro, Phys. Rev. A \textbf{77}, 043809 (2008);
\emph{ibid}. \textbf{79}, 023833.

\bibitem {gat4} A. Gatti, M.Bache, D. Magatti, E. Brambilla, F. Ferri, and L. A. Lugiato, J. Mod. Opt. \textbf{53}, 739 (2006).
\bibitem {bac}  M. Bache, D. Magatti, F. Ferri, A. Gatti,  E. Brambilla, and L. A. Lugiato, Phys. Rev. A \textbf{73}, 053802 (2006).
\bibitem {cai}  Y. Cai and S. Y. Zhu, Phys. Rev. E \textbf{71}, 056607 (2005).
\bibitem {zha}  Y. H. Zhai, X. H. Chen, D. Zhang, and L. A. Wu, Phys. Rev. A \textbf{72}, 043805 (2005).
\bibitem {scar} G. Scarcelli, V. Berardi, and Y. H. Shih, Phys. Rev. Lett. \textbf{96}, 063602 (2006).
\bibitem {dan}  M. D'Angelo, Y. H. Kim, S. P. Kulik, and Y. H. Shih, Phys. Rev. Lett. \textbf{92}, 233601 (2004).
\bibitem {liu}  H. Liu, X. Shen,  D. M.  Zhu, and S. Han, Phys. Rev. A \textbf{76}, 053808 (2007).
\bibitem {zhan} M. Zhang, Q. Wei, X. Shen, Y. Liu, H. Liu, J. Cheng, and S. Han, Phys. Rev. A \textbf{75}, 021803 (2007).

\bibitem {sca}  G. Scarcelli, V. Berardi, and Y. H. Shih, Appl. Phys. Lett. \textbf{88}, 061106 (2006).
\bibitem {shi}  Y. H. Shih, e-print arXiv:0805.1166.
\bibitem {erk}  B. I. Erkmen and J. H. Shapiro, Phys. Rev. A \textbf{78}, 023835 (2008).

\bibitem {ga}   A. Gatti, M. Bondani, L. A. Lugiato, M. G. A. Paris, and C. Fabre, Phys. Rev. Lett. \textbf{98}, 039301 (2007).
\bibitem {wa}   L. G. Wang, S. Qamar, S. Y. Zhu, and M. S. Zubairy, Phys. Rev. A \textbf{79}, 033835 (2009).
\bibitem {bre}  M. E. Brezinski and B. Liu, Phys. Rev. A \textbf{78} 063824 (2008).
\bibitem {sh}   Y. H. Shih, Front. Phys. China, \textbf{2}, 125 (2007).
\bibitem {sha}  J. H. Shapiro, Phys. Rev. A \textbf{78}, 061802(R) (2008).


\bibitem {ou}   L. H. Ou and L. M. Kuang, J. Phys. B \textbf{40}, 1833 (2007).
\bibitem {ba}   Y. F. Bai and S. S. Han, Phys. Rev. A \textbf{76}, 043828 (2007).
\bibitem {li}   J. B.  Liu and Y. H. Shih, Phys. Rev. A \textbf{79}, 023819 (2009).
\bibitem {dz}   D. Z. Cao, J. Xiong, S. H. Zhang, L. F. Lin,  L. Gao, and K. G. Wang, Appl. Phys. Lett. \textbf{92}, 201102 (2008).
\bibitem {aga}  I. N. Agafonov, M. V. Chekhova, T. Sh. Iskhakov, and A. N. Penin, Phys. Rev. A \textbf{77} 053801 (2008).
\bibitem {ric}  Th. Richter, Phys. Rev. A \textbf{42}, 1817 (1990).
\bibitem {man}  L. Mandel and L. Wolf, \emph{Optical Coherence and Quantum Optics} (Cambridge University Press, Cambridge, 1995) p428.
\bibitem{Cao}  D. Z. Cao, J. Xiong, and K. G. Wang,  Phys. Rev. A \textbf{71}, 013801 (2005);
               D. Z. Cao  and K. G. Wang, \emph{Phys. Lett.} A \textbf{333}, 23 (2004).
\bibitem{Bache}M. Bache, D. Magatti, F. Ferri, A. Gatti, E. Brambilla, and L. A. Lugiato, Phys. Rev. A \textbf{73}, 053802 (2006).
\end{thebibliography}
\end{document}